\documentclass[useAMS,usenatbib]{mn2e}
\usepackage{graphicx}

\begin{document}

\title[Correcting systematic effects in photometry]{Correcting 
systematic effects in a large set of photometric lightcurves}

\author[O. Tamuz, T. Mazeh \& S. Zucker]{O.~Tamuz,$^1$\thanks{E-mail:
omert@wise.tau.ac.il} T.~Mazeh$^1$ and S.~Zucker$^{2,3}$\\ $^1$School of
Physics and Astronomy, Raymond and Beverly Sackler Faculty of Exact
Sciences, Tel Aviv University, Tel Aviv, Israel\\ $^2$Observatoire de
Gen\`{e}ve, 51 Ch. des Maillettes, Sauverny CH-1290, Switzerland
\\ $^3$Present address: Faculty of Physics, Weizmann Institute of Science, 
Rehovot 76100, Israel}

\maketitle



\begin{abstract}
We suggest a new algorithm to remove systematic effects in a large set
of lightcurves obtained by a photometric survey. The algorithm can remove 
systematic effects, like the ones associated with atmospheric extinction, 
detector efficiency, or PSF changes over the detector. 
The algorithm works without any prior knowledge of the effects, 
as long as they linearly appear in many stars of the sample. 
The approach, which was originally developed to remove atmospheric extinction 
effects, is based on a lower rank approximation of matrices, an approach 
which was already suggested and used in chemometrics, for 
example. The proposed algorithm is specially useful in cases where 
the uncertainties of the measurements are unequal. For equal uncertainties
the algorithm reduces to the Principal Components Analysis (PCA)
algorithm. We present a simulation to demonstrate the effectiveness of
the proposed algorithm and point out its potential, in search for
transit candidates in particular.
\end{abstract}
\begin{keywords}
atmospheric effects -- 
methods: data analysis --
methods: statistical --
techniques: photometric --
surveys --
planetary systems
\end{keywords}

\section{Introduction}

The advent of large high-$S/N$ CCDs for the use of astronomical
studies has driven many photometric projects that have already produced
unprecedentedly large sets of accurate stellar lightcurves for various
astronomical studies. An example of such a project is the OGLE search for
transit candidates, which has already yielded significant results
\citep[e.g.,][]{Udaetal2002}. Searching for low-amplitude
variables, such as planetary transits, involves finding a faint signal
in noisy data. It is therefore of prime interest to remove any
systematic effects hiding in the data.

Systematic observational effects may be associated, for example, 
with the varying atmospheric conditions, the variability of the detector
efficiency or PSF changes. However, these effects might vary from star
to star, depending on the stellar colour or the position of the star
on the CCD, a dependence which is not always known. Therefore, the
removal of such effects might not be trivial. 

We present here an algorithm to remove some of the systematic
effects in a large set of lightcurves, without any {\it a priori}
knowledge of the different observational features that might affect
the measurements. The algorithm finds the
systematics and their manifestation in the individual stars, as
long as these effects appear in many lightcurves.

We started the development of our algorithm in an attempt to correct
for the atmospheric extinction, with an approach similar to that
of \citet{Semeniuk2003}. We derived the best-fitting airmasses
of the different images and the extinction coefficients of the
different stars, without having any information on the stellar
colours.  However, the result is a general algorithm that deals with
linear systematic effects. It turned out that such an
algorithm had already been proposed by \citet{GabZam1979}, who had applied
it to data from disciplines other than astronomy, chemometrics, for
example. In some restricted cases, when one can ignore the different
uncertainties of the data points, this algorithm reduces to the
well-known Principal Component Analysis
\citep[PCA;][Ch. 2]{MurHec1987}. However, when the uncertainties of the
measurements vary substantially, like in many photometric surveys, PCA
performs poorly, as we demonstrate below.

Section \ref{airmasscolor} presents the initial, simpler version of
our algorithm, which was meant solely to remove the colour-dependent
atmospheric extinction. Section \ref{generalize} puts the algorithm in
a broader context, and shows how the algorithm can remove linear
systematic effects, and can even treat
several unknown effects. Section \ref{simulations} presents
a simulation to demonstrate the effectiveness of our algorithm. We
discuss some of the algorithm properties and potential developments in 
Section \ref{discussion}.

\section{CORRECTION FOR ATMOSPHERIC EXTINCTION}
\label{airmasscolor}

The colour-dependent atmospheric extinction is an obvious
observational effect that contaminates ground-based photometric
measurements.  This effect depends on the stellar
colours, which are not always completely known. This is
specially true for photometric surveys when only one filter is used
and no explicit colour information is available. This section
describes how we find the best stellar extinction coefficients to
account for the atmospheric absorption, together with the most
suitable airmasses assigned for each image.

Consider a set of $N$ lightcurves, each of which consists of $M$
measurements. We define the residual of each observation, $r_{ij}$, to
be the average-subtracted stellar magnitude, i.e., the stellar
magnitude after subtracting the average magnitude of the individual
star.

Let $\{a_j\ ;j=1,...,M\}$ be the airmass at which the $j$-th image was
observed. We can then define the effective extinction coefficient
$c_{i}$ of star $i$ to be the slope of the best linear fit for the
residuals of this star -- $\{r_{ij} \ ;j=1,...,M\}$ as a function of
the corresponding airmasses -- $\{a_j\ ;j=1,...,M\}$. We aim to
remove the product $c_ia_j$ from each $r_{ij}$. In fact, we search for
the best $c_i$ that minimizes the expression
\begin{equation}
S^2_i=\sum_{j}{{\Bigl(r_{ij}-c_ia_j\Bigr)^2}\over{\sigma^2_{ij}}} \ \ ,
\label{eq:S2i}
\end{equation}
where $\sigma_{ij}$ is the uncertainty of the measurement of star $i$
in the image $j$.

Assuming the airmasses are known, a simple differentiation and
equating to zero yields an estimate for the extinction coefficient:
\begin{equation}
c_i={{\sum_{j}{{r_{ij}a_j}\over{\sigma^2_{ij}}}} \over
{\sum_{j}{{a_j^2}\over{\sigma^2_{ij}}}}} \ \ .  \label{eq:iter1}
\end{equation}
Note that the derivation of each $c_i$ is
independent of all the other $c_i$'s, but does depend on all the $\{a_j\}$.

The problem can now be turned around. Since atmospheric extinction
might depend not only on the airmass but also on weather conditions,
we can ask ourselves what is the most suitable "airmass" of each
image, given the known coefficient of every star. Thus we can look for the
$a_j$ that minimizes
\begin{equation}
 S^2_j=\sum_{i}{{\Bigl(r_{ij}-c_ia_j\Bigr)^2}\over{\sigma^2_{ij}}} \ \ , 
\end{equation}
given the previously calculated set
of $\{c_i\}$. The value of the effective 'airmass' is then:
\begin{equation}
a_{j}^{(1)}={{\sum_{i}{{r_{ij}c_i}\over{\sigma^2_{ij}}}}
    \over
    {\sum_{i}{{c_j^2}\over{\sigma^2_{ij}}}}}  \ \ .   \label{eq:iter2}
\end{equation}
We can now recalculate the best-fitting coefficients, $c_{i}^{(1)}$,
and continue iteratively. We thus have an iterative process which in
essence searches for the two sets -- $\{\bar c_i\}$ and $\{\bar
a_j\}$, that best account for the atmospheric extinction.

We performed many simulations that have shown that this iterative
process converged to the same $\{\bar a_j\}$ and $\{\bar c_i\}$, no
matter what initial values were used. Therefore, we suggest that the
proposed algorithm can find the most suitable airmass of each image
and the extinction coefficient of each star. As 
the next section shows, these airmasses and coefficients may have no
relation to actual airmass and colour.

\section{Generalization}
\label{generalize}

The algorithm presented in the previous section is in fact a search to find the 
best two sets of $\{c_i\ ;i=1,N\}$ and $\{a_j\ ;j=1,M\}$  that minimize the 
{\it global} expression
\begin{equation}
S^2=\sum_{ij}{{\Bigl(r_{ij}-c_ia_j\Bigr)^2}\over{\sigma^2_{ij}}} \ \ .
\label{eq:S2}
\end{equation}
Therefore, although the alternating 'criss-cross' iteration process
\citep{GabZam1979} started with the actual airmasses of the different
images, the values of the final set of parameters $\{\bar a_j\}$ and
$\{\bar c_i\}$ are not necessarily related to the true airmass and
extinction coefficient. They are merely the variables by which the
global sum of residuals, $S^2$, varies linearly most significantly.
They could represent any strong systematic effect that might be
associated, for example, with time, temperature or position on the
CCD. This algorithm finds the systematic effect as long as the global
minimum of $S^2$ is achieved.

Now, suppose the data includes a few different systematic effects,
with different $\{c_i\}$ and $\{a_j\}$. We can easily generalize the
algorithm to treat such a case. To do that we denote by $\{^{(1)}\bar
c_i\}$ and $\{^{(1)} \bar a_j\}$ the first set of parameters found in
the data.  We then remove this effect and denote the new residuals by

\begin{equation}
^{(1)}r_{ij}=r_{ij}-{^{(1)}\bar c_i}\  {^{(1)} \bar a_j}\ \ .
\end{equation}

We can then proceed and search for the next linear effect, hidden in
$\{^{(1)}r_{ij}\}$. We use the same procedure to find now the
$^{(2)}c_i$ and $^{(2)}a_j$ that minimize

\begin{equation}
^{(1)}S^{2}=\sum_{ij}{ \bigl({^{(1)} r_{ij}-{^{(2)}c_i}\ {^{(2)} a_j}}\bigr)^2
\over{\sigma^2_{ij}}  }\ \ .
\end{equation}
This process can be applied repeatedly, until it finds no
{\it significant} linear effects in the residuals. The algorithm finds
any linear systematic effect that can be presented as $c_i a_j$ for
the $j$-th measurement of the $i$-th star. 

After developing our algorithm we found that such an approach had
already been proposed by \citet{GabZam1979} as a lower-rank approximation
to data matrices. They applied the algorithm to data from other
disciplines, like climate statistics and chemometrics, and discussed
its convergence properties.  Very similar algorithms were developed
and applied for signal and image processing
\citep[e.g.,][]{Luetal1997}.  If all measurements have the same
uncertainties, the algorithm will reduce to the conventional PCA that can
be applied through the SVD technique \citep[][Ch. 2]{NR}. However, when the uncertainties
of the measurements are substantially different, PCA becomes less
effective at finding and removing systematic effects.  This can lead
to removal of true variability, and can leave some
 actual systematic effects in the data.

\section{Simulation}
\label{simulations}

To demonstrate the power of the algorithm we present here one out of
the many simulations we ran. In this specific example we simulated
lightcurves of 3000 stars in 1000 images. All stars were set to have
constant magnitudes with normally distributed noise of various
amplitudes. We added three systematic effects that depended on
airmass, the CCD position and lunar phase. Finally, we added transit-like
lightcurves to three stars.

To simulate a realistic set of lightcurves, we assigned different noise
levels to different stars, as if we had bright (high $S/N$) and
faint (low $S/N$) objects. The r.m.s.\ ranges between $0.01$ and $1$
mag, with an average value of $0.3$~mag. We assigned to each
measurement an uncertainty which depended on the stellar standard
deviation. To avoid an unrealistic case where all
measurements of a star have the same uncertainty, we randomly varied
the uncertainty of each measurement by 10 percent of the
stellar r.m.s.

We selected three light curves with $0.01$~mag r.m.s and added to them
planetary transit-like signals. The transit periods were $2.7183$, $3.1415$
and $1.4142$ days, with depths of $15$, $20$ and $25$~mmag, respectively.

\begin{figure*}
 \includegraphics{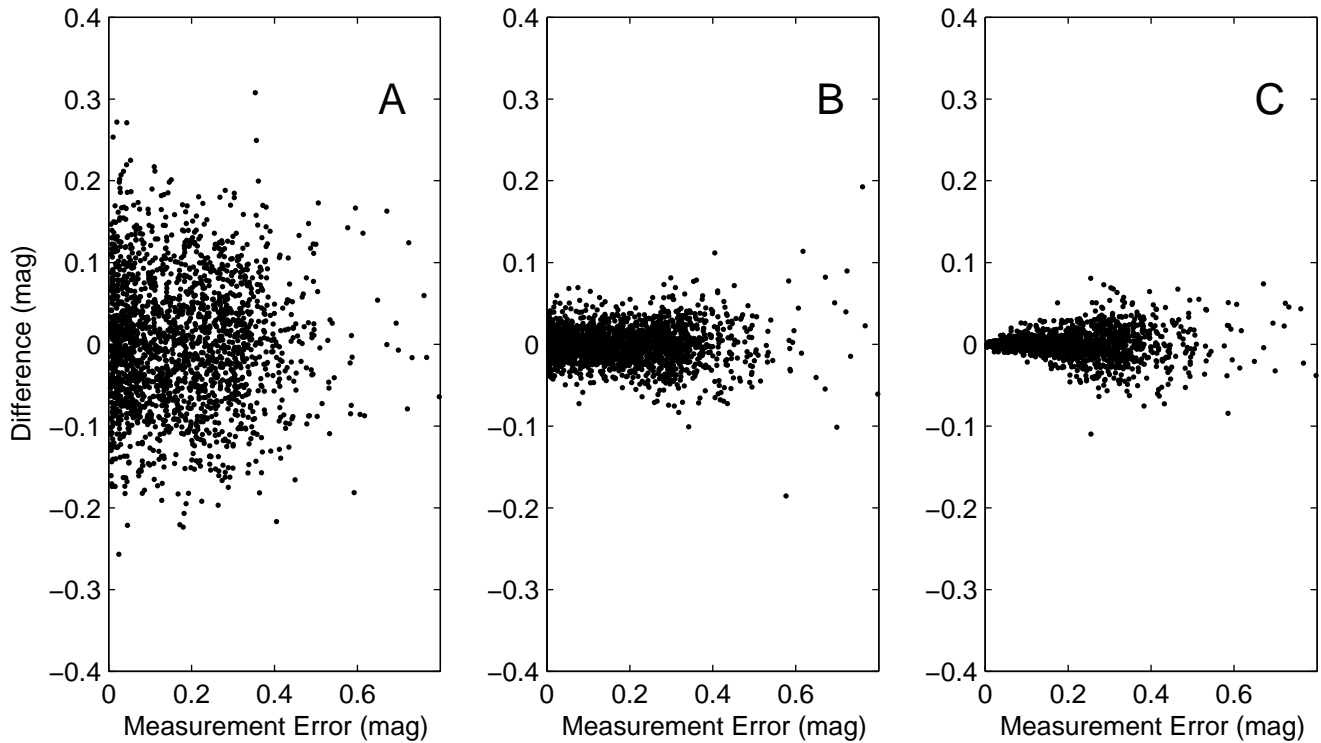} \caption{ 
The differences between
 the derived measurements and their original values (without the
 systematic effects).  Panel (a) presents the measurements before any
 correction was applied, (b) after applying PCA (c) and after applying
 the proposed algorithm (c).  Data is plotted as a function of the
 stellar r.m.s. 2000 randomly chosen measurements are plotted.  
 }
\label{fig:improve}
\end{figure*}

We added three systematic effects that depended 
on airmass (linear and quadratic effects), on the CCD $X$-position and on 
the lunar phase. Thus,
\begin{equation}
r_{ij}=c_i(am)_j+d_i(am)_j^2+x_ib_j+f_i\sin(\omega_{lunar}t_j)+Poisson{\
}noise{\ } [+Transit]\,
\label{eq:noise}
\end{equation} 
where:
\begin{itemize}
\item The observation times $\{t_j\}$ were set to the times of the
  first 1000 images of the OGLE Carina field survey, available from
  the OGLE website\footnote{http://siruis.astrouw.edu.pl/\~{}ogle}.
  
\item The airmasses $\{(am)_j\}$ were calculated using these times and
 the OGLE Carina survey parameters.

\item The positions on the CCD $\{x_i\}$ were randomly drawn from
 a uniform distribution, between $0$ and $2047$.

\item The coefficients $\{c_i\}$ $\{d_i\},\{b_j\}$ and $\{f_i\}$
 were randomly drawn from normal distributions of zero mean. The
 standard deviations of these distributions were chosen so that the
 four systematic effects produced r.m.s.\ variability of $0.06$,
 $0.04$, $0.01$ and $0.008$, respectively.

\end{itemize}

We applied our algorithm to the simulated artificial survey four successive
times, to eliminate four different linear effects (see section
\ref{discussion} for a discussion of the number of effects to
subtract). For comparison, we applied the PCA subtraction to the same
data set the same number of times.

\begin{figure*}
 \includegraphics{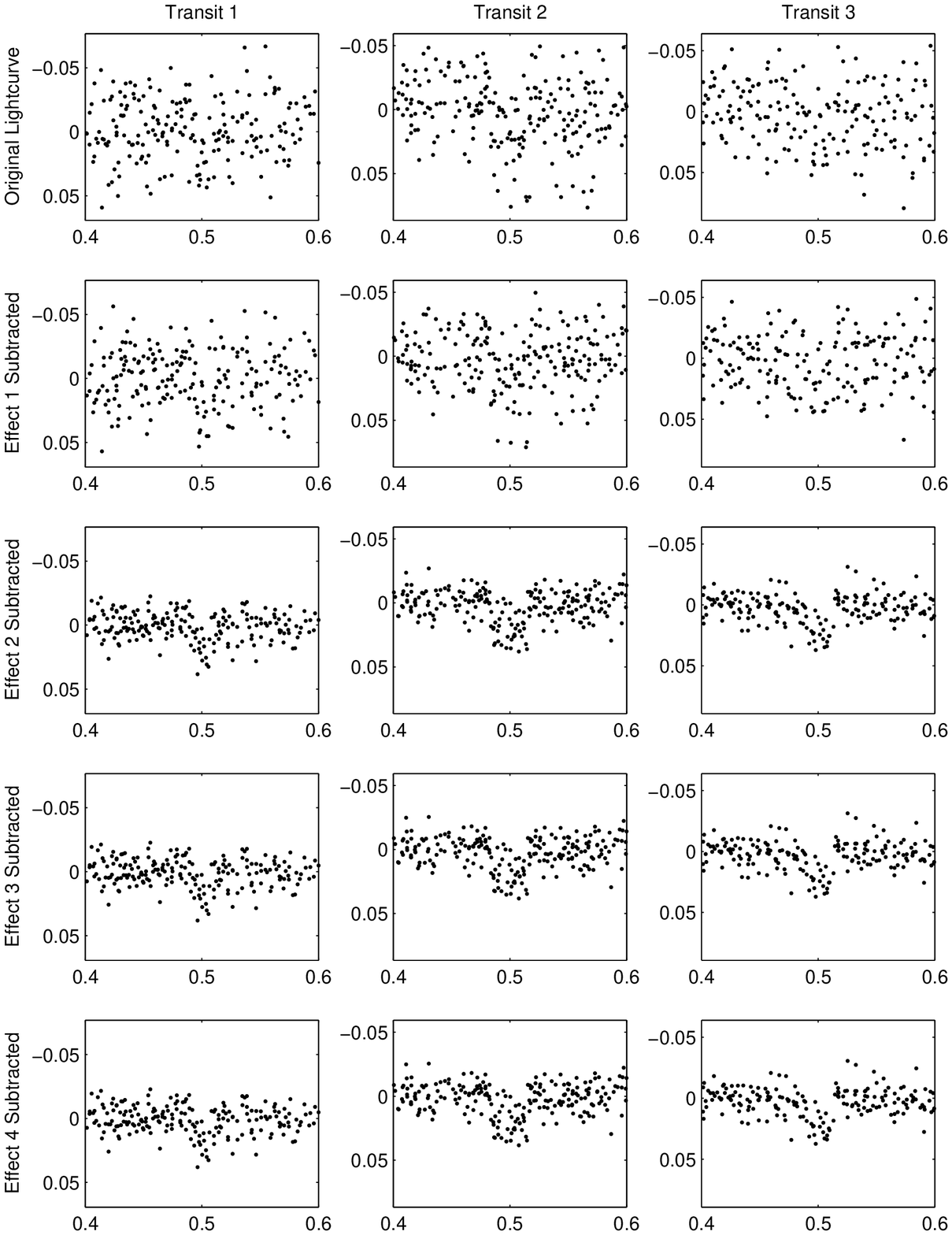} 
 \caption{ 
  Folded lightcurves of the three transits planted in the data 
  before and after the few first iterations.
Each column depicts the lightcurves of one transit. 
The top row shows the data
 uncorrected, while the following rows show it after successive applications
 of the algorithm. The lightcurves are folded on the transit periods
 and show the points which lay within 0.1 phases of the middle of the
 transit.  
}
 \label{fig:transits}
\end{figure*}

The efficacy of the algorithm is demonstrated in two figures.  Fig.\
\ref{fig:improve} presents randomly selected 2000 measurements, before
and after the two algorithms were applied. Panel (A) shows the 
difference between the magnitudes before and after the systematic
effects were added. This difference, which is actually the exact amount
added by the systematic effects, is plotted as a function of the
r.m.s.\ of each star.  We see that the typical systematic error is
about $0.1$--$0.2$~mag. For the ``faint'' stars, with r.m.s.\ of about
$0.4$--$0.7$~mag, the systematic error is relatively small. However,
for the ``bright'' stars in the sample, with inherent r.m.s.\ smaller
than, say, $0.05$ mag, the additional systematic error is relatively
large. This added noise can seriously hamper the ability to detect
small effects like planetary transits.

Panel (B) shows the systematic effects left after PCA 
was applied four times, again as a function
of the stellar r.m.s. Had PCA worked perfectly, all
differences would have been nullified, and all points in the diagram
would have been concentrated on the horizontal line that goes through
zero. We can see that PCA managed to correct all the systematic
effects larger than $0.05$~mag, but failed to perform for smaller
systematic errors.

Panel (C) of Fig.\ \ref{fig:improve} presents the same 2000
measurements, this time after applying our algorithm four times. We
see that the ability of the algorithm to remove the systematic error
depends strongly on the stellar inherent r.m.s. The algorithm performs substantially
better when the stellar r.m.s.\ is small.  For those stars, the
advantage of our algorithm over the PCA approach seems clear. In fact,
all natural candidates for transit detection are exactly those stars.

The ability to detect transits is depicted in Fig.\
\ref{fig:transits}, where we focus on the three stars with simulated
transits. Each column presents the stellar folded lightcurve before
and after the successive iterations were applied. The data were folded
with the transit period, and were plotted around the mid-transit
phase. While initially the systematic errors completely masked the
transits, the three of them gradually surfaced as more iterations were
applied.

\section{Discussion}
\label{discussion}

The proposed algorithm reduces to the PCA approach for the 
case of equal uncertainties.  It is therefore
suggestive to explore the features of our algorithm by analogy with
the corresponding features of PCA.

In case of equal uncertainties the vectors $\{a_j\ ;j=1,...,M\}$ are the 
eigenvectors of the covariance
matrix $R^{\mathrm T}R$, where $R$ is the measurement matrix 
$\{r_{ij}\ ;i=1,...,N,j=1,...,M\}$. Since $R^{\mathrm T}R$ is symmetric,
$\{a_j\ ;j=1,...M\}$ constitute an orthogonal set of vectors. The
first few $a_j$'s are therefore an orthogonal base that spans the
vector subspace of the significant systematic effects. Thus, it may
very well happen that the strongest effect PCA derives is a {\it
linear combination} of some effects we know from prior physical
insight, like a certain combination of the airmass and the $X$
position on the CCD chip. Conversely, it may so happen that two
effects about which we have some insight, such as the $Y$ position and
the lunar phase, span a vector subspace which includes much of the
power of a third effect, say the airmass. In this case, PCA derives
only two significant effects, contrary to our prior intuition.

We suggest that our algorithm exhibits similar behaviour. It is
true that in the general case of unequal uncertainties the
orthogonality of the $\{a_j\}$ is not guaranteed, but the same
qualitative behaviour probably persists.

The recent large photometric surveys and the planned photometric space
missions (e.g., {\it CoRoT}, {\it Kepler}) will face not only the
problem of systematic effects, but also the problem of long-term
stellar variability. It turns out that the proposed solution can
potentially remove some of this variability. In this case,
the various $\{a_j\}$ would assume the values of some function, $f$, of the
timing of the $j$-th image: $a_j=f(t_j)$.  For
equal uncertainties, the space of possible time
variability can be spanned by an orthogonal basis of functions (e.g.,
trigonometric functions in the case of evenly spaced time
sampling). From the PCA point of view, these basis functions, 
(like $\{^{(k)}a_j = \cos(\omega_{(k)} t_j)\ ;k=k_1,k_2\ldots\}$) 
can be thought of as systematic effects. 
The contribution of each basis function to the individual stars is reflected 
through the stellar coefficients, $^{(k)}c_i$. Removing those
effects amounts to removing part of the power of the long-term
variability. Once again, the general case probably shows similar
behaviour.

In general, the main use of PCA has been to reduce the
dimensionality of the data by finding only the significant
factors. Thus, an important question in PCA \citep[][ch. 2]{MurHec1987}
is the number of significant factors to retain. In PCA, it is easy to
solve for the complete set of effects (all eigenvectors of the
covariance matrix), and then decide about the significant factors. In
the general case of unequal uncertainties, we can proceed in two
alternative ways. One way is to solve simultaneously for an assumed
number of effects. The other alternative is to solve for the effects
in sequential stages. In each stage we subtract the effects found in
previous stages before solving for a new effect. The two alternatives
are equivalent in the PCA case (equal uncertainties), but in the
general case they lead to different solutions. Moreover, subtracting
the effect in each stage opens up the possibility of subtracting
the effect not globally, but only from the stars which are most
affected by it. We plan to further explore these issues in order to
gain more insight into the features of the solution.

We are currently applying the algorithm presented here to parts of the
OGLE III data set. We have already found a few intriguing new planetary
transit candidates, and we are still evaluating the statistical
significance of these findings. It would be of great interest to apply
our algorithm to space mission data, like HST photometry, to find out
how large the systematic effects hidden in the data are. As we have
demonstrated, the advantages of our algorithm are most pronounced 
in a data set of high $S/N$ measurements with substantially varying
uncertainties. Data from space missions exactly fit this description.

\section*{Acknowledgments}

This work was supported by the Israeli Science Foundation through
grant no. 03/233. S.Z.\ wishes to acknowledge support by the European
RTN ``The Origin of Planetary Systems'' (PLANETS, contract number
HPRN-CT-2002-00308) in the form of a fellowship.

\bibliographystyle{mn2e}
\bibliography{ref}


\end{document}